\documentclass{article}

\usepackage[a4paper,margin=1in]{geometry}

\usepackage{graphicx,latexsym,amssymb,amsmath}
\usepackage{diagrams}

\newtheorem{definition}{Definition}[section]
\newtheorem{theorem}{Theorem}[section]
\newtheorem{proposition}{Proposition}[section]

\newenvironment{example}{\noindent{\em Example:}}{}

\newcommand{\proof}[1]{{{\em Proof.} #1}}

\def\calF{{\cal F}}
\def\calG{{\cal G}}
\def\calH{{\cal H}}
\def\calS{{\cal S}}

\def\da{\downarrow}
\def\ra{\rightarrow}
\def\fa{\forall}
\newbox\tempa
\newbox\tempb
\newdimen\tempc
\def\mud#1{\hfil $\displaystyle{\mathstrut #1}$\hfil}
\def\rig#1{\hfil $\displaystyle{#1}$}
\def\irulehelp#1#2#3{\setbox\tempa=\hbox{$\displaystyle{\mathstrut #2}$}%
                        \setbox\tempb=\vbox{\halign{##\cr
        \mud{#1}\cr
        \noalign{\vskip\the\lineskip}%
        \noalign{\hrule height 0pt}%
        \rig{\vbox to 0pt{\vss\hbox to 0pt{${\; #3}$\hss}\vss}}\cr
        \noalign{\hrule}%
        \noalign{\vskip\the\lineskip}%
        \mud{\copy\tempa}\cr}}%
                      \tempc=\wd\tempb
                      \advance\tempc by \wd\tempa
                      \divide\tempc by 2 }
\def\irule#1#2#3{{\irulehelp{#1}{#2}{#3}%
                     \hbox to \wd\tempa{\hss \box\tempb \hss}}}

\begin{document}

\title{Enumerating proofs of positive formulae}
\author{Gilles Dowek\thanks{\'Ecole polytechnique and INRIA, LIX, \'Ecole
  polytechnique, 91128 Palaiseau Cedex, France. {\tt gilles.dowek@polytechnique.fr}} and Ying Jiang\thanks{State Key
Laboratory of Computer Science, Institute of Software, Chinese
Academy of Sciences,  Beijing 100190, P.R.China. {\tt jy@ios.ac.cn}}}
\date{}
\maketitle
\thispagestyle{empty}

\begin{abstract}
We provide a semi-grammatical description of the set of normal
proofs of positive formulae in minimal predicate logic, {\em i.e.} a
grammar that generates a set of {\em schemes}, from each of which we
can produce a finite number of normal proofs. This method is
complete in the sense that each normal proof-term of the formula is
produced by some scheme generated by the grammar.  As a corollary,
we get a similar description of the set of normal proofs of positive
formulae for a large class of theories including simple type theory
and System F.
\end{abstract}
\maketitle

\section{Introduction}

A simple way to establish that provability in a logic is decidable
is to develop a proof-search method, enumerating all the potential
proofs of a given formula, and to prove that the search tree of this
method is finite. In this case, when a formula is provable, we can
even conclude that it has a finite number of proofs.  This is
typically the situation in some formulations of classical
propositional sequent calculus \cite{Kleene}.

In some other cases, typically in some formulations of
intuitionistic or minimal propositional sequent calculus, the search
tree is infinite but regular, i.e. it has only a finite number of
distinct sub-trees \cite{Kleene}. In such a situation, provability
is still decidable, but the sets of proofs may be infinite.
Nevertheless, we can describe it with a context-free grammar.

In contrast to Kleene's result, Zaionc has proved that the set of
normal proof-terms of a given formula in minimal propositional logic
({\em i.e.} the set of normal terms of a given type in simply typed
lambda-calculus) is not a context-free language \cite{Zaionc2005}.
This result is a consequence of the undecidability of definability
in simply typed lambda-calculus \cite{Loader} (see also \cite{Joly}
for a minimal example) and it explains why previous grammatical
descriptions of the set of normal terms of a given type had required
an infinite number of symbols
\cite{TakahashiAkamaHirokawa,BenYelles,Hindley,Zaionc1988}.

The reason of this discrepancy between Kleene's and Zaionc's results
is that the former applies to a notion of sequent whose left hand
side is a set and the latter to one whose left hand side is a list.
When using sets, there is no way to distinguish proof-terms such as
$\lambda \alpha:P~\lambda \beta:P~\alpha$ and $\lambda
\alpha:P~\lambda \beta:P~\beta$. These two proof-terms should be
written in the same way using the schematic notation $\lambda
\alpha:P~\lambda \alpha:P~\alpha$.

Using this idea, Takahashi, Akama, and Hirokawa
\cite{TakahashiAkamaHirokawa} as well as Broda and Damas
\cite{BrodaDamas,BrodaDamas2} have shown that if we use such a
schematic language for proof-terms, where identical hypotheses are
referred to by the same name, the set of proof-terms of a given
formula in minimal propositional logic becomes a context-free
language.  Moreover, each schematic proof-term of this context-free
language corresponds to a finite number of genuine proof-terms. For
instance, the schematic proof-term $\lambda \alpha : P~\lambda
\alpha : P~\alpha$ corresponds to two proof-terms: $\lambda \alpha :
P~\lambda \beta : P~\alpha$ and $\lambda \alpha : P~\lambda \beta :
P~\beta$. More generally, each variable occurrence of a schematic
proof-term may be replaced by a variable chosen in a finite set,
yielding a finite number of proof-terms.

\begin{figure*}
\framebox[\textwidth][l]{
\begin{minipage}[l]{\textwidth}
$$\irule{\Delta,
\alpha : A_1 \ra ... \ra A_{n} \ra P \vdash t_1 :
A_1~~~...~~~\Delta, \alpha : A_1 \ra ... \ra A_n \ra P \vdash t_n :
A_n}
        {\Delta, \alpha : A_1 \ra ... \ra A_n \ra P
\vdash (\alpha~t_1~...~t_n):P}
        {\mbox{\em $L\ra$}}$$

if $P$ is atomic.
$$\irule{\Delta \vdash t:A}
        {\Delta \vdash \lambda x~t:\forall x~A}{\mbox{\em $R\forall$}}$$
if $x$ is not free in $\Delta$.
$$\irule{\Delta, \alpha : A \vdash t:B}
        {\Delta \vdash \lambda \alpha~t:A \ra B}{\mbox{\em
        $R\ra$}}$$
\caption{The system LJ$^{+}$: a sequent calculus for positive
sequents}\label{fig:LJ+}
\end{minipage}
}
\end{figure*}

When such a grammar exists, we say that we have a {\em
semi-grammatical description} of the set of proof-terms of a given
formula.  More precisely, a semi-grammatical description of a set is
formed with a context-free grammar and an algorithm generating a
finite number of elements of the set from each element of the
language defined by the grammar.

In \cite{bracket}, we have given a new decidability proof for the
fragment of minimal predicate logic where all quantifiers are
positive and obtained, as a corollary, the decidability of type
inhabitation for positive types in System F. The motivation for
studying the positive fragment of minimal logic is twofold. First,
in the classical case, it is well-known that the undecidability
comes from the negative quantifiers and that the positive fragment
is decidable. The positive fragment, both for classical and minimal
predicate logics, appears to be a large natural decidable fragment.
Secondly, in System F, the datatypes are expressed as positive
types. For instance, the type of unary natural numbers is encoded as
$\fa X~(X \ra (X \ra X) \ra X)$ and that of binary numbers as $\fa
X~(X \ra (X \ra X) \ra (X \ra X) \ra X)$. However, some positive
types, such as $\fa X~(X \ra ((X \ra X) \ra X) \ra X)$, are not
datatypes. Nevertheless, we may want to describe the sets of normal
terms of such types, because they are used in higher-order abstract
syntax or as the input type of the algorithm, extracted from the
constructive proof of the completeness theorem \cite{Krivine}.

The algorithm defined in \cite{bracket} consists in building a
regular search tree, based on a careful handling of variable names
with a system of brackets.  In this paper, we extend the result and
give a semi-grammatical description for the set of $\beta$-normal
$\eta$-long proof-terms of a given formula in the positive fragment
of minimal predicate logic.

First, as the search-tree introduced in \cite{bracket} is regular,
we can define a grammar enumerating the schematic proof-terms. Then,
we give an algorithm to generate a finite set of terms corresponding
to a given scheme. This algorithm is more complex than that for the
propositional case, because the types may be modified when a
variable is replaced by another.  The method obtained in this way is
complete in the sense that each normal proof-term of the formula is
produced from some scheme generated by the grammar. Finally, this
semi-grammatical description of normal proof-terms of positive
formulae also applies to several theories such as simple type theory
and System F.

\section{The systems LJ$^{+}$ and LJB}

\begin{figure*}
\framebox[\textwidth][l]{
\begin{minipage}[l]{\textwidth}
$$\irule{\Gamma^* {\da} \vdash A_1 ~~~ ...  ~~~\Gamma^*{\da} \vdash A_{n}}
        {\Gamma \vdash P} {\mbox{\em $L\ra$}}$$
where\\
$\Gamma = \Gamma_1, [\Gamma_{2}, [... \Gamma_{i-1}, [\Gamma_{i}, A_1
\ra ... \ra A_{n} \ra P]_{V_{i-1}} ...]_{V_{2}}]_{V_1}$,\\
$\Gamma^* = ([...[[\Gamma_1]_{V_1}, \Gamma_{2}]_{V_{2}}, ...
\Gamma_{i-1}]_{V_{i-1}}, \Gamma_{i}, A_1 \ra ... \ra A_{n} \ra P)$, \\
and $P$ is atomic and has no free variable in $V_1 \cup V_{2} \cup
... \cup V_{i-1}$.
$$\irule{[\Gamma]_V {\da} \vdash A}
        {\Gamma \vdash \forall x~A}
        {\mbox{\em $R\forall$}}$$
        where $V$ is the set of all variables bound in $\forall
        x~A$.
$$\irule{(\Gamma, A) {\da} \vdash B}
        {\Gamma \vdash A \ra B}
        {\mbox{\em $R\ra$}}$$
\caption{The system LJB: a sequent calculus with
brackets}\label{fig:LJB}
\end{minipage}}
\end{figure*}

Leaving a more complete description to \cite{bracket}, we briefly
recall, in this section, the notion of positive formula, the sequent
calculi LJ$^{+}$ and LJB. We also introduce a notion of proof-term
to represent derivations in each of these calculi. The proof-terms
of LJ$^+$ are usual lambda-terms and are just called {\em
proof-terms}, while the proof-terms of LJB are called {\em schemes}.

\subsection{Positive formulae}

{\em Minimal predicate logic} is the fragment of predicate logic
with a single connector $\ra$ and a single quantifier $\fa$. Terms
and formulas are defined as usual. A {\em context} is a finite
multiset of formulae and a {\em sequent} $\Gamma \vdash A$ is a pair
formed with a context $\Gamma$ and a formula $A$.

A formula in minimal predicate logic is said to be positive if all
its universal quantifier occurrences are positive. More precisely,
the set of positive and negative formulae and positive sequents in
minimal predicate logic are defined by induction as follows.

\begin{definition}{\bf (Positive and negative formulae)}
\begin{itemize}
\item An atomic formula is {\em positive} and {\em negative}.
\item A formula of the form $A \ra B$ is {\em positive}
(resp. {\em negative}) if $A$ is {\em negative} (resp. {\em positive})
and $B$ is {\em positive} (resp. {\em negative}).

\item A formula of the form $\forall x~A$ is {\em positive} if $A$ is
{\em positive}.
\end{itemize}
\end{definition}

As pointed out in \cite{bracket}, a negative formula has the form
$A_1 \ra ... \ra A_{n} \ra P$, where $P$ is an atomic formula and
$A_1, ..., A_{n}$ are positive formulae.

\begin{definition}{\bf (Positive sequents)}
A sequent $A_1, ..., A_{n} \vdash B$ is {\em positive} if
$A_1$, ..., $A_{n}$ are {\em negative} and $B$ is {\em positive}.
\end{definition}

\subsection{LJ$^{+}$: a sequent calculus for positive sequents}

We use a cut-free sequent calculus for positive sequents in minimal
predicate logic. This sequent calculus contains the usual right rule
for the universal quantifier, but no left rule for this quantifier
is needed because all sequents are positive. It contains also the
usual right rule for the implication. But the left rule for
implication
$$\irule{\Delta, A \ra B \vdash A~~~\Delta, A \ra B, B
\vdash C}
        {\Delta, A \ra B \vdash C}
        {}$$
and the axiom rule
$$\irule{}
        {\Delta, A \vdash A}
        {}$$
are replaced by a more restricted, but equivalent, rule
{\footnotesize
$$\irule{\Delta, A_1 \ra ... \ra A_{n} \ra P \vdash
A_1~~...~~\Delta, A_1 \ra ... \ra A_{n} \ra P \vdash A_{n}}
        {\Delta, A_1 \ra ... \ra  A_{n} \ra
P \vdash P}
        {}$$}
where $P$ is an atomic formula.

In order to associate lambda-terms to proofs, we must associate
proof variables to formulae in contexts. A {\em context with named
formulae} is a finite multiset of pairs, each of them formed with a
{\em proof variable} and a formula, in such a way that each proof
variables occurs at most once. A {\em sequent with named formulae}
$\Delta \vdash A$ is a pair formed with a context $\Delta$ with
named formulae and a formula $A$. These proof variables are
distinguished from the usual term variables of predicate logic.

The rules of the system LJ$^{+}$, equipped with proof-terms, are
depicted in Figure \ref{fig:LJ+}. Notice that all these proof-terms
are $\beta$-normal $\eta$-long. Ignoring these proof-terms, it
yields the original presentation of LJ$^{+}$ given in
\cite{bracket}. When $\Delta \vdash t:A$ is derivable, we also say
that $t$ is a proof-term of the sequent $\Delta \vdash A$.

\subsection{LJB: a sequent calculus with brackets}
\label{sectionLJB}

\begin{figure*}
\framebox[\textwidth][l]{
\begin{minipage}[l]{\textwidth}
{\footnotesize \hspace*{4.1cm}$\irule{\irule{\irule{\irule{\irule{}
                                           {B \ra Q, P(y) \ra Q, P(y) \vdash P(y)}
                                           {L\ra}
                                           \hspace{1cm} ||
                                           \hspace{1.5cm}
                                     \irule{\irule{\irule{
\irule{}{B \ra Q, [P(y) \ra Q, P(y)]_{y}, P(y) \ra Q, P(y) \vdash
P(y)}{L \ra}
                                           \hspace{1cm} ||
                                           \hspace{1cm}
\irule{B \ra Q, [P(y) \ra Q, P(y)]_{y} \vdash (P(y) \ra Q) \ra P(y)
\ra Q}
                                                                {B \ra Q, [P(y) \ra Q, P(y)]_{y}, P(y) \ra Q, P(y) \vdash B}
                                                                {R \fa}
                                                          }
                                                         {B \ra Q, [P(y) \ra Q,P(y)]_{y}, P(y) \ra Q, P(y) \vdash Q}
                                                         {L\ra}
                                                   }
                                                  {B \ra Q, [P(y) \ra Q, P(y)]_{y} \vdash (P(y) \ra Q) \ra P(y) \ra Q}
                                                  {R\ra}
                                            }
                                           {B \ra Q, P(y) \ra Q, P(y) \vdash B}
                                           {R\fa}
                                     }
                                     {B \ra Q, P(y) \ra Q, P(y) \vdash Q}
                                     {L \ra}
                               }
                               {B \ra Q  \vdash B}
                               {R\ra, R\fa}
                       }
                       {B \ra Q \vdash Q}
                       {L\ra}
                }
                {\vdash (B \ra Q ) \ra Q}
                {R\ra}$}
\caption{An example of search tree in LJB.}\label{fig:exampleLJB}
\end{minipage}}
\end{figure*}

\begin{figure*}
\framebox[\textwidth][l]{
\begin{minipage}[l]{\textwidth}
$$\irule{\Gamma^*{\da} \vdash \pi_1:A_1 ~~~ ...  ~~~\Gamma^*{\da} \vdash \pi_n:A_{n}}
        {\Gamma \vdash (\alpha~\pi_1~...~\pi_n):P} {\mbox{\em $L\ra$}}$$
where\\
 $\Gamma = \Gamma_1, [\Gamma_{2}, [... \Gamma_{i-1},
[\Gamma_{i}, A_1 \ra ... \ra A_{n} \ra P]_{V_{i-1}}
...]_{V_{2}}]_{V_1}$,\\
$\Gamma^* = ([...[[\Gamma_1]_{V_1}, \Gamma_{2}]_{V_{2}}, ...
\Gamma_{i-1}]_{V_{i-1}}, \Gamma_{i}, A_1 \ra ... \ra A_{n} \ra
P)$,\\
$P$ is atomic and has no free variable in $V_1 \cup V_{2} \cup ...
\cup V_{i-1}$,\\ and $\alpha$ is the canonical variable of type $A_1
\ra ... \ra A_{n} \ra P$.

$$\irule{[\Gamma]_V {\da}\ \vdash \pi:A}
        {\Gamma \vdash \lambda x~\pi:\forall x~A}
        {\mbox{\em $R\forall$}}$$
        where $V$ is the set of all variables bound in $\forall
        x~A$.

$$\irule{(\Gamma, A){\da} \vdash \pi:B}
        {\Gamma \vdash \lambda \alpha:A~\pi:A \ra B}{\mbox{\em $R\ra$}}$$
where $\alpha$ is the canonical variable of type $A$. \caption{The
system LJB with schemes.}\label{fig:schemes}
\end{minipage}}
\end{figure*}

Search trees in LJ$^{+}$ are not always finite or even regular. For
instance, the search tree of the formula $((P \ra Q) \ra Q) \ra Q$
is infinite and that of the formula $((\fa x~(P(x) \ra Q)) \ra Q)
\ra Q$ is not even regular. To prove the decidability of the
positive fragment of minimal predicate logic, we have introduced in
\cite{bracket} another sequent calculus called LJB.

In LJ$^{+}$, to apply the R$\fa$ rule to the sequent $\Gamma \vdash
\forall x~A$, we have to rename the variable $x$ either in $\fa x~A$
or in $\Gamma$ so that the variable released by the rule does not
appear in the context. In LJB, instead of renaming the variable $x$,
we bind it in the context $\Gamma$ with brackets and obtain the
sequent $[\Gamma]_{x} \vdash A$. In fact, for technical reasons, we
bind in $\Gamma$, not only the variable $x$, but also all the bound
variables of $A$.

\begin{definition}{\bf (LJB-contexts and items)}
{\em LJB-contexts} and {\em items} are mutually inductively
defined as follows.
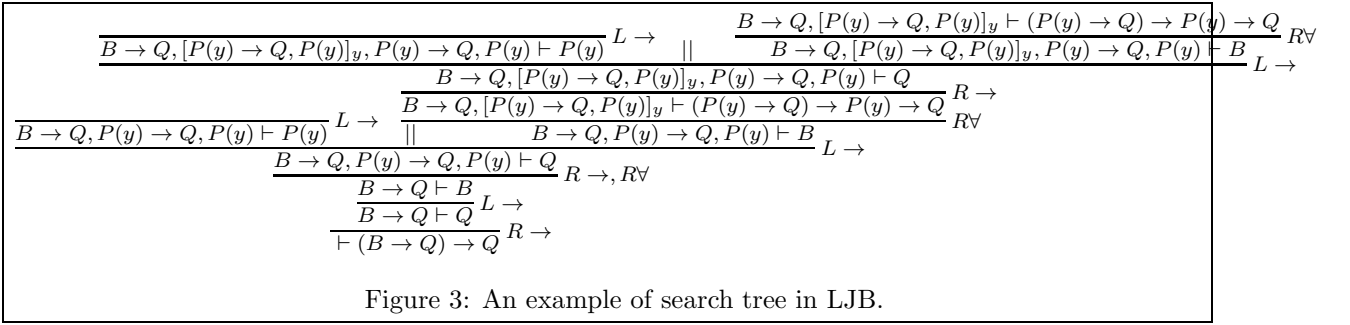
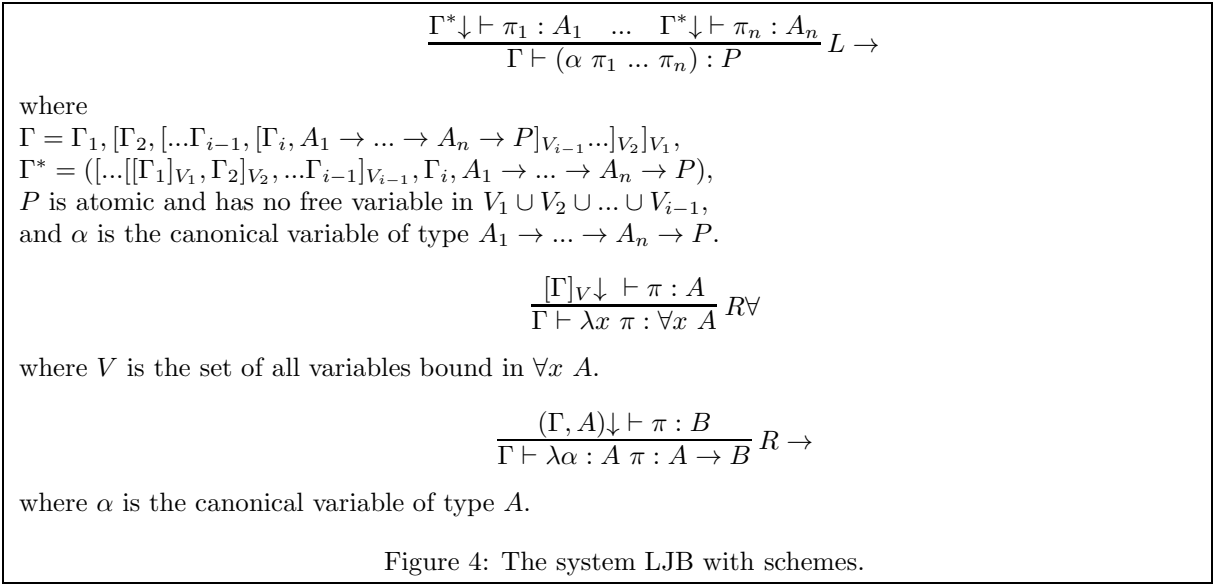
\begin{itemize}
\item A {\em LJB-context} $\Gamma$ is a finite multiset of items
$\{I_1, ..., I_{n}\}$.

\item An {\em item} $I$ is either a formula or an expression of
the form $[\Gamma]_V$ where $V$ is a set of variables and $\Gamma$ a
LJB-context.
\end{itemize}
\end{definition}

In the item $[\Gamma]_V$, the variables of $V$ are bound by the
symbol $[~]$.

A {\em LJB-sequent} $\Gamma \vdash A$ is a pair formed by a
LJB-context $\Gamma$ and a formula $A$.

The system LJB is formed by two sets of rules: the usual deduction
rules and additional transformation rules dealing with bracket
manipulation. The transformation rules form a terminating rewrite
system: the first rule allows to replace an item of the form $[I,
\Gamma]_V$ by the two items $I$ and $[\Gamma]_V$ provided no free
variable of $I$ is in $V$; the second one allows to remove trivial
items; the third rule to replace two identical items by one.

\begin{definition}{\bf (Cleaning LJB-contexts)}
\label{cleaning} The {\em cleaning rules} are
$$\begin{array}{ll}
[I, \Gamma]_V \longrightarrow I, [\Gamma]_V, & \mbox{if
\(FV(I) \cap V = \varnothing\)}\\
\left[\ \right]_V \longrightarrow \varnothing &\\
I I \longrightarrow I&
\end{array}$$
where $I$ is an item and $\Gamma$ a LJB-context.
\end{definition}

Instead of proving the confluence of the rewrite system of
Definition \ref{cleaning}, we fix an arbitrary strategy and define
the normal form $\Gamma{\da}$ of a context $\Gamma$ as the normal
form relative to this strategy. We may, for instance, proceed as
follows. If $\Gamma = \varnothing$ then we let $\Gamma{\da} =
\varnothing$. Otherwise, we choose an item $I$ in $\Gamma$ and let
$\Gamma' = \Gamma \setminus \{I\}$. Then, we normalize the item $I$
and the LJB-context $\Gamma'$ recursively. We let $\Gamma{\da} =
\Gamma'{\da}$ if $I{\da}$ is an element of $\Gamma'{\da}$ and
$\Gamma{\da} = I{\da}, \Gamma'{\da}$ otherwise. To normalize an item
$I$, we need to consider the two following cases. If $I$ is a
formula, then we let $I{\da} = I$. If it has the form $[\Delta]_V$,
we first normalize recursively $\Delta$, then we let $\Delta_1$ be
the part of $\Delta{\da}$ formed with the elements that have a free
variable in $V$ and let $\Delta_2 = \Delta{\da} \setminus \Delta_1$.
Finally, we let $I{\da} = \Delta_2$ if $[\Delta_1]_V$ is an element
of $\Delta_2$ and $I{\da} = [\Delta_1]_V, \Delta_2$ otherwise.

The deduction rules apply to LJB-sequents with normalized contexts
with respect to the cleanning rules and where the bound variables
are named differently and differently from the free variables. It is
easy to check that these properties are preserved by the rules.
Moreover, in LJB we deal with formulae, not formulae modulo
$\alpha$-equivalence.

The rules of the system LJB are depicted in Figure \ref{fig:LJB}. In
the $L\ra$ rule, brackets are moved from some items of the
LJB-context to others, bringing the formula $A_1 \ra ... \ra A_{n}
\ra P$ inside brackets to the surface, so that it can be used. For
instance the LJB-sequent $Q(x), [Q(x) \ra P]_{x} \vdash P$ is
transformed (bottom-up) into $[Q(x)]_{x}, Q(x) \ra P \vdash Q(x)$.
The crucial point is that the two occurrences of $x$ in $Q(x)$ and
$Q(x) \ra P$ that are separated in the first LJB-sequent remain
separated.

The main interest of the system LJB is that, as illustrated in the
Example \ref{exampleLJB}, the search tree in LJB of any positive
formula is regular. This property is a consequence of the following
proposition proved in \cite{bracket} (Proposition 4.5).

\begin{proposition}
Let $A$ be a positive formula. There exists a finite set $\calS$ of
sequents such that all the sequents occurring in a LJB-proof of the
sequent $\vdash A$ are in $\calS$.
\end{proposition}

\begin{example}\label{exampleLJB}
Let $A = (B \ra Q) \ra Q$ where $B = \fa y~((P(y) \ra Q) \ra (P(y)
\ra Q))$. The search tree of the sequent $\vdash A$ is given in
Figure \ref{fig:exampleLJB}.

Notice that when trying to prove the sequent $B \ra Q, P(y) \ra Q,
P(y) \vdash Q$ we may apply the $L\ra$ rule either with the
proposition $B \ra Q$ or with the proposition $P(y) \ra Q$, yielding
two branches in the search tree. The same holds with the sequent $B
\ra Q, [P(y) \ra Q,P(y)]_{y}, P(y) \ra Q, P(y) \vdash Q$. Notice
also that the search tree is infinite and regular. We have cut the
infinite branch when the sequent $B \ra Q, [P(y) \ra Q, P(y)]_{y}
\vdash (P(y) \ra Q) \ra P(y) \ra Q$ appeared for the second time.
\end{example}

\subsection{Schemes}

Now we introduce {\em schemes}, that are the proof-terms for the
system LJB. Unlike what we did for LJ$^{+}$, we do not assign names
to hypotheses in LJB. Instead, we choose a canonical proof variable
for each such formula. The rules of LJB with schemes are depicted in
Figure \ref{fig:schemes}.

\begin{figure*} \framebox[\textwidth][l]{
\begin{minipage}[l]{\textwidth}
$$s_{\Gamma \vdash P}  \longrightarrow
(\alpha~ s_{\Gamma^*{\da} \vdash A_1}~...~s_{\Gamma^*{\da} \vdash
A_n})$$ where\\
$\Gamma = \Gamma_1, [\Gamma_{2}, [... \Gamma_{i-1}, [\Gamma_{i}, A_1
\ra ... \ra A_{n} \ra P]_{V_{i-1}} ...]_{V_{2}}]_{V_1}$,\\
$\Gamma^* = ([...[[\Gamma_1]_{V_1}, \Gamma_{2}]_{V_{2}}, ...
\Gamma_{i-1}]_{V_{i-1}}, \Gamma_{i}, A_1 \ra ... \ra A_{n} \ra
P)$,\\
$P$ is atomic and has no free variable in $V_1 \cup V_{2} \cup ...
\cup V_{i-1}$,\\ and $\alpha$ is the canonical variable of type $A_1
\ra ... \ra A_n \ra P$.

$$s_{\Gamma \vdash \fa x~A} \longrightarrow \lambda x~
s_{[\Gamma]_V{\da} \vdash A}$$ where $V$ is the set of all variables
bound in $\forall x~A$.

$$s_{\Gamma \vdash A \ra B} \longrightarrow \lambda \alpha:A~s_{(\Gamma,
A){\da}  \vdash B}$$ where $\alpha$ is the canonical variable of
type $A$. \caption{The scheme grammar.} \label{fig:grammar}
\end{minipage}}
\end{figure*}

\section{A grammar to enumerate schemes}

In this section, we prove that, although it may be infinite, the set
of schemes of a given normalized LJB-sequent may be described by a
context-free grammar.

\begin{definition}[Scheme grammar]
\label{grammar} Let $\Gamma \vdash A$ be a normalized LJB-sequent
and $\calS$ be the finite set of sequents that may occur in a
derivation of $\Gamma \vdash A$.  To each sequent $S$ of $\calS$, we
associate a non-terminal symbol $s_{S}$ and set up the rules
displayed in Figure \ref{fig:grammar}.
\end{definition}

The grammar generating the schemes of the type $A$ given in Example
\ref{exampleLJB} and a scheme generated by this grammar are detailed
in the example below.

\begin{example}
The grammar generating the schemes of the type $A = (B \ra Q) \ra Q$
where $B = \fa y~((P(y) \ra Q) \ra (P(y) \ra Q))$ is
$$
\begin{array}{lll}
S & \ra & \lambda \alpha~(\alpha~\lambda y\lambda \beta\lambda
\gamma~(\beta~\gamma))\\
S & \ra & \lambda \alpha~(\alpha~\lambda y\lambda \beta\lambda
\gamma~(\alpha~\lambda y~S_1))\\
S_1 & \ra & \lambda \beta\lambda \gamma~(\beta~\gamma)\\
S_1 & \ra & \lambda \beta\lambda \gamma~(\alpha~\lambda y~S_1)
\end{array}$$
where $S$ is the non-terminal associated to the sequent $\vdash A$,
$S_1$ that associated to $B \ra Q, [P(y) \ra Q, P(y)]_{y} \vdash
(P(y) \ra Q) \ra P(y) \ra Q$, $\alpha$ is the canonical variable of
type $B \ra Q$, $\beta$ that of type $P(y) \ra Q$ and $\gamma$ that
of type $P(y)$.

A scheme generated by the grammar is
$$\lambda \alpha~(\alpha~\lambda y\lambda \beta~\lambda
\gamma~(\alpha~\lambda y\lambda \beta~\lambda
\gamma~(\beta~\gamma)))$$
\end{example}

\begin{proposition}[Soundness]
\label{SoundGramm} Let $\Gamma \vdash A$ be a normalized
LJB-sequent. Then for any scheme $\pi$ generated in $s_{\Gamma
\vdash A}$, we have $\Gamma \vdash \pi:A$.
\end{proposition}

\proof{By induction on the derivation of $\pi$ in the grammar.
\hfill$\Box$}

\begin{proposition}[Completeness]
\label{CompGramm} Let $\Gamma \vdash A$ be a normalized LJB-sequent.
Then each scheme $\pi$ such that $\Gamma \vdash \pi:A$ is generated
in $s_{\Gamma \vdash A}$.
\end{proposition}

\proof{By induction on the derivation of $\Gamma \vdash \pi : A$ in
the system LJB with schemes.\hfill$\Box$}

\section{Generating proof-terms}

Now we are ready to provide a term enumeration algorithm through the
grammatical scheme enumeration algorithm described in the previous
section. In this endeavor, we will define a function $\calH$, which,
roughly speaking, associates a finite set of terms to a scheme, in
such a way that $t$ is a proof-term if and only if there exists a
scheme $\pi$ such that $t \in \calH(\pi)$. To define this function
$\calH$, we need a function ${\cal G}$ handling context cleaning.
When defining the function $\calG$, the only non trivial case is
that of the rule $II \longrightarrow I$, which is handled in turn by
another function $\calF$.

Definitions \ref{s1} and \ref{s2} below extend the usual notion of
$\alpha$-equivalence for formulae to sequents of LJ$^{+}$ and LJB,
and will be useful in the rest of the section.

\begin{definition}[$\alpha$-equivalence of sequents]\label{s1}
Two sequents $\Gamma \vdash A$ and $\Gamma' \vdash A'$ are said to
be $\alpha$-equivalent if there exists a variable renaming $\sigma$
of term variables (i.e. an injective substitution mapping variables
to variables) such that $\Gamma'$ is $\alpha$-equivalent to $\sigma
\Gamma$ and $A'$ is $\alpha$-equivalent to $\sigma A$.
\end{definition}

For instance, the sequents $P(x) \vdash P(x)$ and $P(y) \vdash P(y)$
are $\alpha$-equivalent. The intuition is that the variables free in
$\Gamma$ and $A$ are considered as implicitly bound by the symbol
$\vdash$ in the sequent $\Gamma \vdash A$.

We also extend the notion of $\alpha$-equivalence  to sequents of
LJ$^{+}$ with named formulae as follows.

\begin{definition}[$\alpha$-equivalence of sequents with named formulae]\label{s2}
Two sequents $\Gamma \vdash A$ and $\Gamma' \vdash A'$ are said to
be $\alpha$-equivalent if there exists a variable renaming $\sigma$
of term and proof variables such that $\Gamma'$ is
$\alpha$-equivalent to $\sigma \Gamma$ and $A'$ is
$\alpha$-equivalent to $\sigma A$.
\end{definition}

For instance, the sequents $\alpha:P(x) \vdash P(x)$ and $\beta:P(y)
\vdash P(y)$ are $\alpha$-equivalent.

\begin{definition}[Fresh $\alpha$-variant and flattening]\label{ff}
Let $\Gamma \vdash A$ be a normalized LJB-sequent, a {\em fresh
$\alpha$-variant} $\Gamma' \vdash A'$ of $\Gamma \vdash A$ is an
LJB-sequent, which is $\alpha$-equivalent to $\Gamma \vdash A$ and
where all bound variables are named differently.

A LJ$^{+}$-sequent $\Delta \vdash B$ is said to be a {\em
flattening} of a normalized LJB-sequent $\Gamma \vdash A$, if it is
obtained by erasing all the brackets in a fresh $\alpha$-variant of
$\Gamma \vdash A$ and naming all the formulae in $\Gamma$ with
distinct proof variables.
\end{definition}

\begin{example}
A flattening of the LJB-sequent $[P(x), P(x) \ra Q]_x, [P(x), P(x)
\ra Q]_x \vdash Q$ is the LJ$^{+}$-sequent $\alpha_1:P(x_1),
\beta_1:(P(x_1) \ra Q), \alpha_{2}:P(x_{2}), \beta_{2}:(P(x_{2}) \ra
Q) \vdash Q$.
\end{example}

Remark that two flattenings of the same LJB-sequent are
$\alpha$-equivalent LJ$^{+}$-sequents.

\begin{definition}[Partial duplication]
Let $\Sigma \vdash A$ be a sequent of LJ$^{+}$. A sequent $\Delta
\vdash B$ of LJ$^{+}$ is said to be a {\em partial duplication} of
$\Sigma \vdash A$ if there exist two substitutions $\sigma_1$ and
$\sigma_2$ of term-variables with the same domain, renaming the
variables of their domain with fresh and distinct variables such
that for each variable $\gamma:C$ of $\Sigma$, $\Delta$ contains
either the variable $\gamma_1:\sigma_1 C$ or the variable
$\gamma_2:\sigma_2 C$ or both, and $B$ is either $\sigma_1 A$ or
$\sigma_2 A$.
\end{definition}

\begin{example}
If the sequent $\Sigma \vdash A$ is
$$\alpha:(Px \ra Q), \beta : Px \vdash Q$$
and $\sigma_1 = \sigma_2 = id$, then one partial duplication is the
sequent
$$\alpha_1:(Px \ra Q), \beta_1 : Px, \alpha_2:(Px \ra Q),
\beta_2 : Px \vdash Q$$
If the sequent $\Sigma \vdash A$ is
$$\alpha:(Px \ra Q), \beta : Px \vdash Q$$
but $\sigma_1 = x_1/x$ and $\sigma_2 = x_2/x$, then one partial
duplication is the sequent
$$\alpha_1:(Px_1 \ra Q), \beta_1 : Px_1, \alpha_2:(Px_2 \ra Q),
\beta_2 : Px_2 \vdash Q$$
If the sequent $\Sigma \vdash A$ is
$$\alpha:(Px \ra Q), \beta : Px \vdash Px$$
and $\sigma_1 = x_1/x$ and $\sigma_2 = x_2/x$, then one partial
duplication is the sequent
$$\alpha_1:(Px_1 \ra Q), \beta_1 : Px_1, \alpha_2:(Px_2 \ra Q),
\beta_2 : Px_2 \vdash Px_1$$
\end{example}

\begin{definition}[The function $\calF$]
\label{defF} Let $\Sigma \vdash A$ be a sequent of LJ$^{+}$ and
$\Delta \vdash B$ a partial duplication of this sequent obtained
with the substitutions $\sigma_1$ and $\sigma_2$.

\begin{diagram}
& & & &  \mbox{LJ$^{+}$} & \\
& & & & \Delta \vdash B & \calF(u)\\
& & & & & \uTo \\
& & & & \Sigma \vdash A  & u\\
\end{diagram}

\smallskip

Let $u$ be a proof-term of $\Sigma \vdash A$. We define, by
induction on the structure of $u$, a finite set $\calF_{\Sigma
\vdash A}^{\Delta \vdash B}(u)$ of proof-terms of $\Delta \vdash B$.

\begin{itemize}
\item If $u = (\alpha~u_1~...~u_n)$, then $A$ is atomic.
Let $C_1 \ra ... \ra C_n \ra A$ be the type of $\alpha$. For $i \in
\{1,2\}$, if $\Delta$ contains a variable $\alpha_i: \sigma_i C_1
\ra ... \ra \sigma_i C_n \ra \sigma_i A$ and $\sigma_i A = B$, then
we take all terms of the form $(\alpha_i~u'_1~...~u'_n)$ where
$u'_1$ is an element of $\calF_{\Sigma \vdash C_1}^{\Delta \vdash
\sigma_i C_1}(u_1)$, ..., $u'_n$ is an element of $\calF_{\Sigma
\vdash C_n}^{\Delta \vdash \sigma_i C_n} (u_n)$, otherwise we take
no term with head variable $\alpha_i$.

\item If $u = \lambda x~u_1$, then $A$ has the form $\fa x~A_1$ and
$B$ has the form $\fa x~B_1$, where $B_1$ is either $\sigma_1 A_1$
or $\sigma_2 A_1$, we take all terms of the form $\lambda x~u'_1$
where $u'_1$ is an element of $\calF_{\Sigma \vdash A_1}^{\Delta
\vdash B_1}(u_1)$.

\item If $u = \lambda \alpha~u_1$, then $A$ has the form $A_1 \ra A_2$
and $B$ has the form $B_1 \ra B_2$, where $B_1$ is either $\sigma_1
A_1$ or $\sigma_2 A_1$ and $B_2$ is either $\sigma_1 A_2$ or
$\sigma_2 A_2$, we take all terms of the form $\lambda \alpha'~u'_1$
with $u'_1$ an element of $\calF_{\Sigma, \alpha:A_1 \vdash
A_2}^{\Delta, \alpha':B_1 \vdash B_2} (u_1)$.
\end{itemize}
\end{definition}

\begin{example}
If the sequent $\Sigma \vdash A$ is
$$\alpha:(Px \ra Q), \beta : Px \vdash Q$$
$\sigma_1 = \sigma_2 = id$ and one partial duplication is the
sequent
$$\alpha_1:(Px \ra Q), \beta_1 : Px, \alpha_2:(Px \ra Q),
\beta_2 : Px \vdash Q$$ then
$$\calF_{\Sigma \vdash Q}^{\Delta \vdash Q}((\alpha~\beta)) =
\{(\alpha_1~\beta_1), (\alpha_1~\beta_2),
(\alpha_2~\beta_1), (\alpha_2~\beta_2)\}$$
If the sequent $\Sigma \vdash A$ is
$$\alpha:(Px \ra Q), \beta : Px \vdash Q$$
$\sigma_1 = x_1/x$ and $\sigma_2 = x_2/x$ and one partial
duplication is the sequent
$$\alpha_1:(Px_1 \ra Q), \beta_1 : Px_1, \alpha_2:(Px_2 \ra Q),
\beta_2 : Px_2 \vdash Q$$ then
$$\calF_{\Sigma \vdash Q}^{\Delta \vdash Q}((\alpha~\beta)) =
\{(\alpha_1~\beta_1), (\alpha_2~\beta_2)\}$$
Notice that, after having chosen $\alpha_1$, in the first
case, we obtain
$$\calF_{\Sigma \vdash Px}^{\Delta \vdash Px}(\beta) = \{\beta_1,
\beta_2\}$$
while in the second, we obtain
$$\calF_{\Sigma \vdash Px}^{\Delta \vdash Px_1}(\beta) = \{\beta_1\}$$
Our relatively liberal notion of partial duplication allows the
``pathological'' example where the set $\calF_{\Sigma \vdash
A}^{\Delta \vdash B}(u)$ is empty: if the sequent $\Sigma \vdash A$
is
$$\alpha:(Px \ra Q), \beta : Px \vdash Q$$
and $\sigma_1 = x_1/x$ and $\sigma_2 = x_2/x$, then one partial
duplication is the sequent
$$\alpha_1:(Px_1 \ra Q), \beta_2 : Px_2 \vdash Q$$
and $\calF_{\Sigma \vdash A}^{\Delta \vdash B}((\alpha~\beta)) =
\varnothing$.
\end{example}

\begin{proposition}[Soundness]\label{SoundF}
Let $\Delta \vdash B$ be a partial duplication of $\Sigma \vdash A$.
If $u$ is a proof of $\Sigma \vdash A$, and $t \in \calF_{\Sigma
\vdash A}^{\Delta \vdash B}(u)$, then $t$ is a proof of $\Delta
\vdash B$.
\end{proposition}

\proof{By induction on the structure of $u$. \hfill$\Box$}

\begin{proposition}[Completeness]\label{CompF}
Let $\Delta \vdash B$ be a partial duplication of $\Sigma \vdash A$.
If $t$ is a proof of $\Delta \vdash B$ then there exists a proof
$u$, of the same height as $t$, of $\Sigma \vdash A$ such that $t
\in \calF_{\Sigma \vdash A}^{\Delta \vdash B}(u)$.
\end{proposition}

\proof{By induction on the structure of $t$. The term $u$ is
obtained by replacing each variable of the form $\sigma_1 x$ or
$\sigma_2 x$ by $x$. \hfill$\Box$}

\begin{definition}[The function $\calG$]\label{defG}
 Let $\Gamma \vdash A$ be a normalized LJB-sequent and
$\Gamma{\da} \vdash A$ its normal form. Let $\Delta \vdash B$ be a
flattening of $\Gamma \vdash A$ and $\Delta' \vdash B'$ a flattening
of $\Gamma{\da} \vdash A$.

\begin{diagram}
 \mbox{LJB} & & & &  \mbox{LJ$^{+}$} & \\
\Gamma \vdash A & &  \rTo^{\mbox{flattening}} & & \Delta \vdash B & \calG(u)\\
\dTo & & & & & \uTo \\
\Gamma{\da} \vdash A & &\rTo^{\mbox{flattening}}& & \Delta' \vdash B'  & u\\
\end{diagram}

\smallskip

For any proof-term $u$ of $\Delta' \vdash B'$, we construct a set
$\calG_{\Delta' \vdash B'}^{\Delta \vdash B}(u)$ of proof-terms of
$\Delta \vdash B$ by induction on the length of the reduction from
$\Gamma$ to $\Gamma{\da}$.

\begin{itemize}
\item If $\Gamma{\da} = \Gamma$, then
$\Delta' \vdash B'$ and $\Delta \vdash B$ are $\alpha$-equivalent,
thus there exists a renaming $\sigma$ of the free variables of
$\Delta$ and $B$ such that $\Delta$ is $\alpha$-equivalent to
$\sigma \Delta'$ and $B$ is $\alpha$-equivalent to $\sigma B'$. We
take $\calG_{\Delta' \vdash B'}^{\Delta \vdash B}(u) = \{\sigma
u\}$.

\item If $\Gamma$ rewrites to $\Gamma_1$ in one cleaning step
and then $\Gamma_1$ rewrites to $\Gamma{\da}$, then let $\Delta_1
\vdash B_1$ be a flattening of $\Gamma_1 \vdash A$ and let $S =
\calG_{\Delta' \vdash B'}^{\Delta_1 \vdash B_1}(u)$. Now consider
the rule used to reduce $\Gamma$ to $\Gamma_1$. If this rule is
$[~]_V \ra \varnothing$ or $[\Gamma, I]_V \ra [\Gamma]_V, I$ then
$\Delta \vdash B$ and $\Delta_1 \vdash B_1$ are $\alpha$-equivalent,
thus there exists a renaming $\sigma$ of the free variables of
$\Delta$ and $B$ such that $\Delta$ is $\alpha$-equivalent to
$\sigma \Delta_{1}$ and $B$ is $\alpha$-equivalent to $\sigma
B_{1}$. We take $\calG_{\Delta' \vdash B'}^{\Delta \vdash B}(u) =
\{\sigma t~|~t \in S\}$. If this rule is $II \ra I$ then $\Delta
\vdash B$ is a partial duplication of $\Delta_1 \vdash B_1$. We take
$\calG_{\Delta' \vdash B'}^{\Delta \vdash B}(u) = \bigcup_{t \in S}
\calF_{\Delta_1 \vdash B_1}^{\Delta \vdash B}(t)$.
\end{itemize}
\end{definition}

\begin{example}
The sequent $$[P(x), P(x) \ra Q]_{x}, [P(x), P(x) \ra Q]_{x} \vdash
Q$$ normalizes to
$$[P(x), P(x) \ra Q]_{x} \vdash Q$$
A flattening of the first sequent is $\Delta \vdash Q$ where
$\Delta$ is the context {\small $$\alpha_1 : P(x_1), \beta_1 :
      (P(x_1) \ra Q), \alpha_{2} : P(x_{2}), \beta_{2} : (P(x_{2}) \ra Q)$$}and a flattening of the second one is the sequent $\Delta' \vdash
      Q$ where $\Delta'$ is the context
$$\alpha : P(x), \beta : (P(x) \ra Q)$$
Then $$\calG_{\Delta' \vdash Q}^{\Delta \vdash Q}((\alpha~\beta)) =
\{(\beta_1~\alpha_1), (\beta_{2}~\alpha_{2})\}$$
\end{example}

\begin{proposition}[Soundness]\label{SoundG}
Let $\Gamma \vdash A$ be a normalized LJB-sequent and $\Gamma{\da}
\vdash A$ its normal form. Let $\Delta \vdash B$ be a flattening of
$\Gamma \vdash A$ and $\Delta' \vdash B'$ a flattening of
$\Gamma{\da} \vdash A$. Let $u$ be a proof-term of $\Delta' \vdash
B'$ and $t \in \calG_{\Delta' \vdash B'}^{\Delta \vdash B}(u)$. Then
$t$ is a proof-term of $\Delta \vdash B$.
\end{proposition}

\proof{By induction on the length of the reduction from $\Gamma$ to
$\Gamma{\da}$, using Proposition \ref{SoundF} for the case of the
rule $II \longrightarrow I$. \hfill$\Box$}

\begin{proposition}[Completeness]\label{CompG}
Let $\Gamma \vdash A$ be a normalized LJB-sequent and $\Gamma{\da}
\vdash A$ its normal form. Let $\Delta \vdash B$ be a flattening of
$\Gamma \vdash A$ and $\Delta' \vdash B'$ a flattening of
$\Gamma{\da} \vdash A$. If $t$ is a proof of $\Delta \vdash B$, then
there exists a proof $u$, of the same height as $t$, of $\Delta'
\vdash B'$ such that $t \in \calG_{\Delta' \vdash B'}^{\Delta \vdash
B}(u)$.
\end{proposition}

\proof{By induction on the length of the reduction from $\Gamma$ to
$\Gamma{\da}$, using Proposition \ref{CompF} for the case of the
rule $II \longrightarrow I$. \hfill$\Box$}

\begin{definition}[The function $\calH$]
\label{defH} Let $\Gamma \vdash A$ be a normalized LJB-sequent and
$\Delta \vdash B$ a flattening of $\Gamma \vdash A$.

\begin{diagram}
 \mbox{LJB} & & & &  \mbox{LJ$^{+}$} \\
\Gamma \vdash A & &  \rTo^{\mbox{flattening}} & & \Delta \vdash B\\
\pi & & \rTo & & \calH(\pi)\\
\end{diagram}

\smallskip

Let $\pi$ be a scheme of the sequent $\Gamma \vdash A$, we associate
to $\pi$ a set $\calH_{\Gamma \vdash A}^{\Delta \vdash B}(\pi)$ of
proof-terms of type $\Delta \vdash B$ in LJ$^{+}$ by induction on
the structure of $\pi$.

\begin{itemize}
\item
If $\pi = (\alpha~\pi_1~\ldots~\pi_{n})$, then let $A_1 \ra \ldots
\ra A_n \ra A$ be the type of $\alpha$. Select the occurrences of
the formula $A_1 \ra \ldots \ra A_n \ra A$ in $\Gamma$, such that
the rule $L\ra$ can be applied to this occurrence, and for all $i$,
the scheme $\pi_i$ has type $\Gamma^*{\da} \vdash A_i$ where
$\Gamma^*{\da}$ is the context obtained by applying $L\ra$ to this
occurrence. For each selected occurrence, let $\alpha':B_1 \ra
\ldots \ra B_n \ra B$ be the corresponding declaration in $\Delta$.
The sequent $\Delta \vdash B$ is also a flattening
of $\Gamma^* \vdash A$ and the sequent $\Delta \vdash B_i$ is one of
$\Gamma^* \vdash A_i$. Consider a flattening $\Delta' \vdash B'_i$
of $\Gamma^*{\da} \vdash A_i$, set up $S_i = \calH_{\Gamma^*{\da}
\vdash A_i}^{\Delta' \vdash B'_i}(\pi_i)$ and $S'_i = \bigcup_{t \in
S_i} \calG_{\Delta' \vdash B'_i}^{\Delta \vdash B_i}(t)$.  The set
$\calH_{\Gamma \vdash A}^{\Delta \vdash B}(\pi)$ contains the terms
of the form $(\alpha'~t_1~...~t_n)$ for some $\alpha':B_1 \ra \ldots
\ra B_n \ra B$ in $\Delta$ corresponding to a selected occurrence
and $t_i \in S'_i$.

\item
If $\pi = \lambda x~\pi_1$, then $A = \fa x~A_1$, $B = \fa y~B_1$
and $\pi_1$ is a scheme of $[\Gamma]_V{\da} \vdash A_1$. The sequent
$\Delta \vdash B_1$ is a flattening of $[\Gamma]_V \vdash A_1$. Let
$\Delta' \vdash B'_1$ be a flattening of $[\Gamma]_V{\da} \vdash
A_1$, set up $S = \calH_{[\Gamma]_V{\da} \vdash A_1}^{\Delta' \vdash
B'_1} (\pi_1)$ and $S'= \bigcup_{t \in S} \calG_{\Delta' \vdash
B'_1}^{\Delta \vdash B_1}(t)$.  The set $\calH_{\Gamma \vdash
A}^{\Delta \vdash B}(\pi)$ is the set of the terms of the form
$\lambda y~t$ for $t$ in $S'$.

\item
If $\pi = \lambda \alpha:A_1~\pi_1$, then $A = A_1 \ra A_2$ and $B =
B_1 \ra B_2$ and $\pi_1$ is a scheme of $(\Gamma, A_1){\da} \vdash
A_2$.  The sequent $\Delta, \alpha':B_1 \vdash B_2$ is a flattening
of $\Gamma, A_1 \vdash A_2$. Let $\Delta' \vdash B'_2$ be a
flattening of $(\Gamma, A_1){\da} \vdash A_2$, set up $S =
\calH_{(\Gamma, A_1){\da} \vdash A_2}^{\Delta' \vdash B'_2}(\pi_1)$
and $S' = \bigcup_{t \in S} \calG_{\Delta' \vdash B'_2}^{\Delta,
\alpha':B_1 \vdash B_2}(t)$.  The set $\calH_{\Gamma \vdash
A}^{\Delta \vdash B}(\pi)$ is the set of the terms of the form
$\lambda \alpha':B_1~t$ for $t$ in $S'$.
\end{itemize}
\end{definition}

\begin{example}
Continuing the Example \ref{exampleLJB}, let
$$
\pi = \lambda \alpha~(\alpha~\lambda y\lambda \beta~\lambda
\gamma~(\alpha~\lambda y\lambda \beta~\lambda
\gamma~(\beta~\gamma)))$$ The set ${\calH}_{\vdash A}^{\vdash
A}(\pi)$ contains the two terms
$$\begin{array}{l}
\lambda \alpha~(\alpha~\lambda y_1\lambda \beta_1\lambda
\gamma_1~(\alpha~\lambda y_2\lambda \beta_2\lambda
\gamma_2~(\beta_1~\gamma_1)))\\
\lambda \alpha~(\alpha~\lambda y_1\lambda \beta_1\lambda
\gamma_1~(\alpha~\lambda y_2\lambda \beta_2\lambda
\gamma_2~(\beta_2~\gamma_2)))
\end{array}$$
where $\alpha : B \ra Q$, $\beta_1 : P(y_1) \ra Q$, $\gamma_1 :
P(y_1)$, $\beta_2 : P(y_2) \ra Q$, $\gamma_2 : P(y_2)$.
\end{example}

\begin{proposition}[Soundness]\label{Sound}
Let $\Gamma \vdash A$ be a normalized LJB-sequent and $\Delta \vdash
B$ be a sequent of LJ$^{+}$ that is a flattening of $\Gamma \vdash
A$. Then for each scheme $\pi$ of $\Gamma \vdash A$, every
proof-term in $\calH_{\Gamma \vdash A}^{\Delta \vdash B}(\pi)$ is a
proof-term of $\Delta \vdash B$.
\end{proposition}

\proof{By induction on the height of $\pi$, using Proposition
\ref{SoundG} for context cleaning.\hfill$\Box$}

\begin{proposition}[Completeness] \label{Complete}
Let $\Gamma \vdash A$ be a normalized LJB-sequent and $\Delta \vdash
B$ a sequent of LJ$^{+}$ such that $\Delta \vdash B$ is a flattening
of $\Gamma \vdash A$. Then for each proof-term $t$ of $\Delta \vdash
B$, there exists a scheme $\pi$ of $\Gamma \vdash A$ such that $t
\in \calH_{\Gamma \vdash
  A}^{\Delta \vdash B}(\pi)$.
\end{proposition}

\proof{By induction on the structure of $t$.
\begin{itemize}
\item If $t = (\alpha'~t_1~...~t_n)$, then the variable $\alpha' : B_1
\ra ... \ra B_n \ra B$ is declared in $\Delta$ and $t_i$ is a
proof-term of $\Delta \vdash B_i$.  The variable $\alpha'$
corresponds to an occurrence of a formula $A_1 \ra ... \ra A_n \ra
A$ in $\Gamma$ and $\Gamma$ has the form $\Gamma_1, [\Gamma_2, [...
\Gamma_{i-1}, [\Gamma_{i}, A_1 \ra ... \ra A_{n} \ra A]_{V_{i-1}}
...]_{V_2}]_{V_1}$.  As $\Delta \vdash B$ is a flattening of $\Gamma
\vdash A$ and this occurrence of $A_1 \ra ... \ra A_{n} \ra A$
corresponds to $B_1 \ra ... \ra B_{n} \ra B$, $A$ has no free
variable in $V_1 \cup V_2 \cup ... \cup V_{i-1}$. Thus, the sequent
$\Delta \vdash B$ is also a flattening of $\Gamma^* \vdash A$ and
$\Delta \vdash B_i$ is a flattening of $\Gamma^* \vdash A_i$.

Let $\Delta' \vdash B'_i$ be a flattening of $\Gamma^{*}{\da} \vdash
A_i$. By Proposition \ref{CompG}, there exists a proof-term $u_i$ of
$\Delta' \vdash B'_i$ of the same height as $t_i$ such that $t_i \in
\calG_{\Delta' \vdash B'_i}^{\Delta \vdash B_i}(u_i)$. By induction
hypothesis, for each $i \in \{1, \ldots, n\}$, there exists scheme
$\pi_i$ of $\Gamma^*{\da} \vdash A_i$ such that $u_i \in
\calH_{\Gamma^*{\da} \vdash A_i} ^{\Delta' \vdash B'_i}(\pi_i)$. So,
if $\alpha$ is the canonical variable of type $A_1 \ra ... \ra A_n
\ra A$, then $(\alpha~\pi_1~...~\pi_n)$ is a scheme of  $\Gamma
\vdash A$ and $(\alpha'~t_1~...~t_n) \in \calH_{\Gamma \vdash
A}^{\Delta \vdash B}(\alpha~\pi_1~...~\pi_n)$.

\item If $t = \lambda y~t_1$, then $B = \forall y~B_1$, $A = \forall
x~A_1$ and $t_1$ is a proof-term of $\Delta \vdash B_1$ that is a
flattening of $[\Gamma]_V \vdash A_1$. Let $\Delta' \vdash B'_1$ be a
flattening of $[\Gamma]_V{\da} \vdash A_1$. By Proposition
\ref{CompG}, there exists a proof-term $u_1$ of $\Delta' \vdash B'_1$ of the
same height as $t_1$ such that $t_1 \in \calG_{\Delta' \vdash
  B'_1}^{\Delta \vdash B_1}(u_1)$. By induction
hypothesis, there exists a scheme $\pi_1$ of $[\Gamma]_V{\da} \vdash
A_1$ such that $u_1 \in \calH_{[\Gamma]_V{\da} \vdash A_1}^ {\Delta'
\vdash B'_1}(\pi_1)$. This implies $\lambda y~t_1 \in \calH_{\Gamma
\vdash A}^{\Delta \vdash B}(\lambda x~\pi_1)$.

\item If $t = \lambda \alpha':B_1~t_1$, then $B = B_1 \ra B_2$,
$A = A_1 \ra A_2$ and $t_1$ is a proof-term of $\Delta, B_1 \vdash
B_2$ that is a flattening of $\Gamma, A_1 \vdash A_2$. Let $\Delta'
\vdash B'_2$ be a flattening of $(\Gamma, A_1){\da} \vdash A_2$. By
Proposition \ref{CompG}, there exists a proof-term $u_1$ of $\Delta'
\vdash B'_2$ of the same height as $t_1$ such that $t_1 \in
\calG_{\Delta' \vdash B'_2}^{\Delta, B_{1} \vdash B_2}(u_1)$. By
induction hypothesis, there exists a scheme $\pi_1$ of $(\Gamma,
A_1){\da} \vdash A_2$ such that $u_1 \in \calH_{(\Gamma, A_1){\da}
\vdash A_2}^{\Delta' \vdash B'_2}(\pi_1)$. Let $\alpha$ be the
canonical variable of type $A_1$, we have $\lambda \alpha'~t_1 \in
\calH_{\Gamma \vdash A}^{\Delta
  \vdash B}(\lambda \alpha~\pi_1)$. \hfill$\Box$
\end{itemize}}

\begin{theorem}
Let $A$ be a formula. Then $t$ is a proof-term of $\vdash A$ in
LJ$^{+}$ if and only if there exists a scheme $\pi$ generated by the
grammar given in Definition \ref{grammar} such that $t \in
\calH_{\vdash A}^{\vdash A}(\pi)$.
\end{theorem}

\proof{From Propositions \ref{SoundGramm}, \ref{CompGramm},
\ref{Sound}, and \ref{Complete}.\hfill$\Box$}

\section{Enumerating normal terms of a positive type in System F}

\begin{figure*}
\framebox[\textwidth][l]{
\begin{minipage}[l]{\textwidth}
$$\irule{\irule{\irule{\irule{\irule{}
                                   {B \ra X, Y \ra X, Y \vdash Y}
                                   {L\ra}
                             \hspace{1cm} || \hspace{1cm}
                             \irule{B \ra X, Y \ra X, Y \vdash X}
                                   {B \ra X, Y \ra X, Y \vdash B}
                                   {R \ra}
                                     }
                                     {B \ra X, Y \ra X, Y \vdash X}
                                     {L \ra}
                               }
                               {B \ra X  \vdash B}
                               {R\ra}
                       }
                       {B \ra X \vdash X}
                       {L\ra}
                }
                {\vdash \fa X \fa Y ((B \ra X) \ra X)}
                {R\ra, R\fa}$$
\caption{A search tree in System F.}\label{fig:example1}
\end{minipage}}
\end{figure*}
As remarked in \cite{bracket}, to each positive type $T$ of System
F, we can associate a formula $\Phi(T)$ in predicate logic with a
single unary predicate $\varepsilon$.
$$\Phi(X) = \varepsilon(X)$$
$$\Phi(T \ra U) = \Phi (T) \ra \Phi (U)$$
$$\Phi(\forall X~T) = \forall X~\Phi(T)$$
and the normal terms of type $T$ in System F are exactly the
proof-terms of $\Phi(T)$ in predicate logic. Thus, the enumeration
algorithm described in the previous sections applies immediately to
System F. The examples below (where we write $X$ for
$\varepsilon(X)$) illustrate the algorithm.

\begin{example}
Let $A = \fa X ((\fa Y((Y \ra X) \ra (Y \ra X)) \ra X ) \ra X)$. Let
$\alpha : \fa Y((Y \ra X) \ra (Y \ra X)) \ra X, \beta : Y \ra X$ and
$\gamma : Y$. Let $S = S_{\vdash A}$ and $S_1 = S_{B \ra X, [Y \ra
X, Y]_{Y} \vdash (Y \ra X) \ra Y \ra X}$. The scheme grammar is
given by
$$\begin{array}{lll}
S & \ra & \lambda X~\lambda \alpha~(\alpha~\lambda Y\lambda
\beta~\lambda
\gamma~(\beta~\gamma))\\
S & \ra & \lambda X~\lambda \alpha~(\alpha~\lambda Y\lambda
\beta~\lambda
\gamma~(\alpha~\lambda Y~S_1))\\
S_1 & \ra & \lambda \beta~\lambda \gamma~(\beta~\gamma)\\
S_1 & \ra & \lambda \beta~\lambda \gamma~(\alpha~\lambda Y S_1)
\end{array}$$
It is easy to check that the scheme below is generated by the
grammar
$$\lambda X~\lambda \alpha~(\alpha~\lambda Y~\lambda \beta~\lambda
\gamma~(\alpha~\lambda Y~\lambda \beta~\lambda
\gamma~(\beta~\gamma)))$$ And this scheme generates in turn two
proof-terms:
$$
\begin{array}{l}
\lambda X~\lambda \alpha~(\alpha~\lambda Y_1~\lambda \beta_1~\lambda
\gamma_1~(\alpha~\lambda Y_2~\lambda
\beta_2~\lambda \gamma_2~(\beta_1~\gamma_1)))\\
\lambda X~\lambda \alpha~(\alpha~\lambda Y_1~\lambda \beta_1~\lambda
\gamma_1~(\alpha~\lambda Y_2~\lambda \beta_2~\lambda
\gamma_2~(\beta_2~\gamma_2)))
\end{array}
$$
where $\alpha : B \ra X$, $\beta_1 : Y_1 \ra X$, $\gamma_1 : Y_1$,
$\beta_2 : Y_2 \ra X$, $\gamma_2 : Y_2$.

More generally, one scheme of depth $n$ generated by this grammar,
yields $n - 1$ proof-terms of type $A$.
\end{example}

\begin{example}
Consider now the prenex form of the formula of the previous example.
Let $A = \fa X \fa Y ((B \ra X ) \ra X)$  where $B = (Y \ra X) \ra
(Y \ra X)$. The search tree of $A$ is given in Figure
\ref{fig:example1}.

Let $\alpha : ((Y \ra X) \ra (Y \ra X)) \ra X, \beta : Y \ra X$ and
$\gamma : Y$. Let $S = S_{\vdash A}$ and $S_1 = S_{B \ra X, Y \ra X,
Y \vdash X}$. The corresponding scheme grammar is given by
$$
\begin{array}{lll}
S & \ra & \lambda X~\lambda Y~\lambda \alpha~(\alpha~\lambda \beta~\lambda \gamma~S_1)\\
S_1 & \ra & (\beta~\gamma)\\
S_1 & \ra & (\alpha~\lambda \beta~\lambda \gamma~S_1)
\end{array}
$$
It is easy to check that the scheme below is generated by the
grammar
$$
\lambda X~\lambda Y~\lambda \alpha~(\alpha~\lambda \beta~\lambda
\gamma~ (\alpha~\lambda \beta~\lambda \gamma~(\beta~\gamma)))$$ And
this scheme generates in turn four proof-terms
$$\begin{array}{l}
\lambda X~\lambda Y~\lambda \alpha~(\alpha~\lambda \beta_1~\lambda \gamma_1~(\alpha~\lambda \beta_2~\lambda \gamma_2~(\beta_1~\gamma_1)))\\
\lambda X~\lambda Y~\lambda \alpha~(\alpha~\lambda \beta_1~\lambda \gamma_1~(\alpha~\lambda \beta_2~\lambda \gamma_2~(\beta_1~\gamma_2)))\\
\lambda X~\lambda Y~\lambda \alpha~(\alpha~\lambda \beta_1~\lambda \gamma_1~(\alpha~\lambda \beta_2~\lambda \gamma_2~(\beta_2~\gamma_1)))\\
\lambda X~\lambda Y~\lambda \alpha~(\alpha~\lambda \beta_1~\lambda
\gamma_1~(\alpha~\lambda \beta_2~\lambda
\gamma_2~(\beta_2~\gamma_2)))
\end{array}$$
where $\alpha : B \ra X$, $\beta_1 : Y \ra X$, $\gamma_1 : Y$,
$\beta_2 : Y \ra X$, $\gamma_2 : Y$.

More generally, one scheme of depth $n$ generated by this grammar,
yields $(n - 1)^2$ proof-terms.
\end{example}

\section*{Conclusion}

Once more, the complexity of predicate logic comes from the negative
quantifiers: when they are removed, not only the logic becomes
decidable, but also the proofs have a simple structure.

The usual interpretations of proofs as terms are based on
formulations of deduction where contexts are multisets or lists. The
schemes are the counterpart to these terms when contexts are sets.
Their structure is even simpler than that of terms and their
interest may go beyond the proof enumeration problem.


\begin{thebibliography}{9}
\bibitem{Kleene}
Kleene, S.C. (1952) Introduction to Metamathematics. North-Holland.

\bibitem{Zaionc2005}
Zaionc, M. (2005) Probabilistic approach to the lambda definability
for fourth order types. Electronic Notes in Theoretical Computer
Science, 140, 41-54.

\bibitem{Loader}
Loader, R. (2001) The undecidability of lambda-definability. Logic,
Meaning and Computation: Essays in Memory of Alonzo Church, Kluwer,
331-342.

\bibitem{Joly}
Joly, Th. (2005) On lambda-definability I: the fixed model problem
and generalizations of the matching problem. Fundam. Inform.,
65(1-2), 135-151.

\bibitem{TakahashiAkamaHirokawa}
Takahashi, M. Akama, Y., and Hirokawa, S. (1996) Normal proofs and
their grammar. Information and Computation, 125(2), 144-153.

\bibitem{BenYelles}
Ben-Yelles, C.B. (1979) Type-assignment in the Lambda-calculus;
Syntax and Semantics. Doctoral Thesis.

\bibitem{Hindley} Hindley, J.R. (1997)
Basic Simple Type Theory. Cambridge Universty Press.

\bibitem{Zaionc1988}
Zaionc, M. (1988) Mechanical procedure for proof construction via
closed terms in typed lambda-calculus. Journal of Automated
Reasoning, 4, 173-190.

\bibitem{BrodaDamas}
Broda, S. and Damas, L. (2001) A context-free grammar representation
for normal inhabitants of types in TA-lambda. EPIA'01, LNAI 2258.

\bibitem{BrodaDamas2}
Broda, S. and Damas, L. (2005) On long normal inhabitants of a type.
J. of Logic and Computation 15, 353-390.

\bibitem{bracket}
Dowek, G. and Jiang, Y. (2006) Eigenvariables, bracketing and the
decidability of positive minimal predicate logic.  Theoretical
Computer Science, 360, 193-208.

\bibitem{Krivine}
Krivine, J.-L. (1996) Une preuve formelle et intuitionniste du
th\'eor\`eme de compl\'etude de la logique classique. Bull. Symb.
Log. 2(4), 405-421.

\end{thebibliography}
\end{document}